\documentclass[final,aps,3p,times,onecolumn]{elsarticle}

\usepackage{amssymb}
\usepackage{amsmath}
\usepackage{latexsym,epsfig}
\usepackage{graphicx}
\usepackage{hyperref}

\usepackage{amsfonts}
\usepackage{epsfig}
\usepackage{dcolumn}
\usepackage{bm}

\def\to{\rightarrow}
\newcommand{\eslash}{\ensuremath{{\hbox{$E_T$\kern-0.9em\lower-.05ex\hbox{/}\kern0.10em\;\;}}}}

\journal{Physics Letters B}

\begin{document}

\begin{frontmatter}



\title{The 750 GeV $S$-cion: Where else should we look for it?}


\author[a1]{Alexandre Alves}
\author[a2]{Alex G. Dias}
\author[a3]{Kuver Sinha}

\address[a1]{Universidade Federal de S\~ao Paulo, Departamento de Ci\^encias Naturais e da Terra,  Diadema - SP, Brasil}
\address[a2]{Universidade Federal do ABC, Centro de Ci\^encias Naturais e Humanas,  Santo Andr\'e - SP, Brasil}
\address[a3]{Department of Physics, Syracuse University, Syracuse, NY 13244, USA}


\begin{abstract}
The resonance $S$ at $\sim 750$ GeV in the diphoton channel observed by ATLAS and CMS, if it holds up, is almost certainly the ($S$)cion of a larger dynasty in a UV completion that may very well be connected to the hierarchy problem. At this stage, however, an effective field theory framework provides a useful way to parametrize searches for this resonance in other channels. Assuming that the excess is due to a new scalar or pseudoscalar boson, we study associated production of $S$ (``$S$-strahlung") at the LHC and propose searches in several clean channels like $\gamma\gamma\ell\ell$, $\gamma\gamma\ell\eslash$ and $\ell\ell\ell\gamma\eslash$ to probe dimension-5 operators coupling $S$ to Standard Model gauge bosons.
We consider a range of widths for $S$, from 5 GeV to 45 GeV, and find that the three channels probe complementary regions of parameter space and the suppression scale $\Lambda$. The finding of most immediate relevance is that with 3 fb$^{-1}$, the LHC might already reveal new excesses in the $\gamma\gamma\ell\ell$ channel and a 5(3) $\sigma$ discovery may already be possible after collecting 65(25) fb$^{-1}$ of data with $\ell\ell\ell\gamma\eslash$ events if the scale of the new physics is within $\sim $ 9 TeV for couplings respecting 8 TeV LHC bounds and compatible with the observed excess in diphotons for a wide resonance as suggested by the ATLAS Collaboration.  Beyond the EFT parametrization, we found realizations of models with heavy vector-like quarks and leptons which can simultaneously fit the diphoton excess and be discovered in the channels proposed here.
\end{abstract}

\begin{keyword}
Diphoton excess\sep scalar resonance\sep effective couplings\sep vector-like quarks and leptons

\end{keyword}

\end{frontmatter}


\section{Introduction}
\label{intro}

ATLAS and CMS recently announced tantalizing hints of a new resonance with mass $\sim 750$ GeV \cite{atlasresult, cmsresult}. The excess is in the diphoton channel, suggesting a spin-$0$ resonance, which is the possibility we will pursue in this paper.

A new scalar or pseudoscalar $S$ (which, for convenience, we call the $S$-cion) would fit in nicely with scenarios beyond the Standard Model (SM), including those which address the hierarchy problem. Several such explanations have already appeared after the Collaborations announced their results \cite{Harigaya:2015ezk}-\cite{Chao:2015nsm}. However, in the absence of more evidence which allows us to lock into a specific model, a pragmatic way to proceed is to use effective field theory (EFT) to parametrize our ignorance. This is the path taken in \cite{Buttazzo:2015txu,Franceschini:2015kwy}, for example. Within such a framework, it becomes possible to investigate the electroweak couplings of $S$ to SM particles and propose searches in various channels.

The purpose of the present work is to point out that additional information can be gathered from the search for signals related to vector boson and $S$-cion associated production (``$S$-strahlung"). Constraints in all possible channels from Run I and the current Run of the LHC are taken into account. We find that searching for the resonance in events containing hard photons, leptons and missing energy in the channels $\ell\ell\gamma\gamma$, $\gamma\gamma\ell\eslash$ and $\ell\ell\ell\gamma\eslash$ might reveal new excesses with the current accumulated data already. Denoting the scale where new physics is introduced in the EFT of $S$ and SM gauge bosons by $\Lambda$, we estimate that the 13 TeV LHC can probe $\Lambda$ up to 9 TeV with 300 fb$^{-1}$ of data in all those channels. With the current integrated luminosity of $\sim 3$ fb$^{-1}$, $\Lambda \sim 3$ TeV may be probed already with a few $\ell\ell\gamma\gamma$ events lurking in the data. 

Our work is model independent in two senses: first, the proposed channels do not involve gluon-fusion, but quark initiated processes through couplings to the SM $Z$, $W$ and photon; second, the scalar couples to the SM gauge bosons via dimension-5 effective operators that respect the SM gauge symmetries. Thus, the results presented in this work would apply to any model where a new 750 GeV scalar couples to SM gauge bosons and fits the diphoton excess.  As a concrete example, we examine models with heavy vector-like quarks and leptons coupling to a CP-even scalar. We found that depending on the fermions multiplicities, electric charges and color representation, signals as the ones proposed in this work can be observed at the LHC from 30 to 3000 fb $^{-1}$, thus confirming the usefulness of the EFT description.



%
\begin{figure}[t]
  \centering
  \includegraphics[width=.45\textwidth]{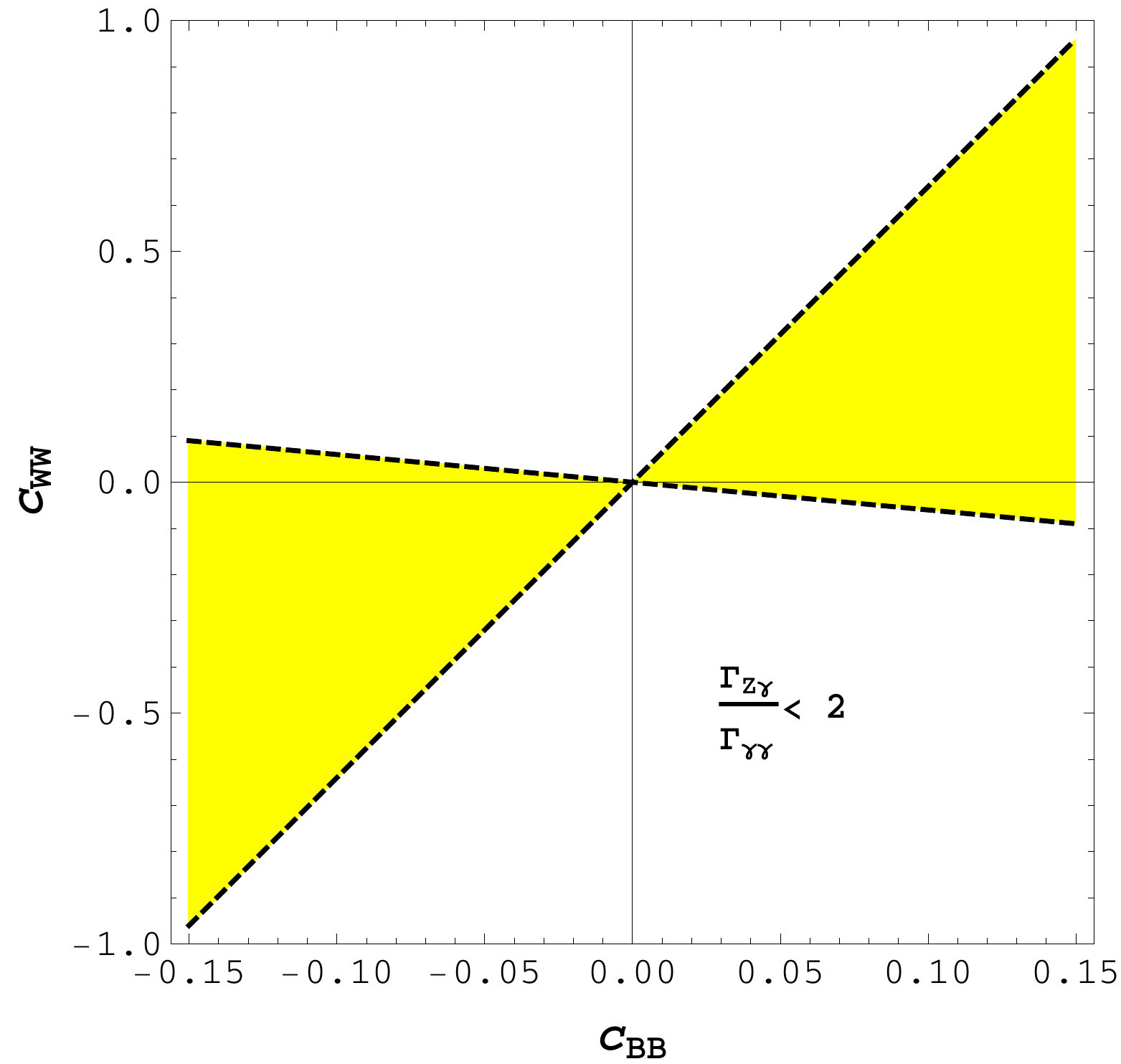}
  \includegraphics[width=.45\textwidth]{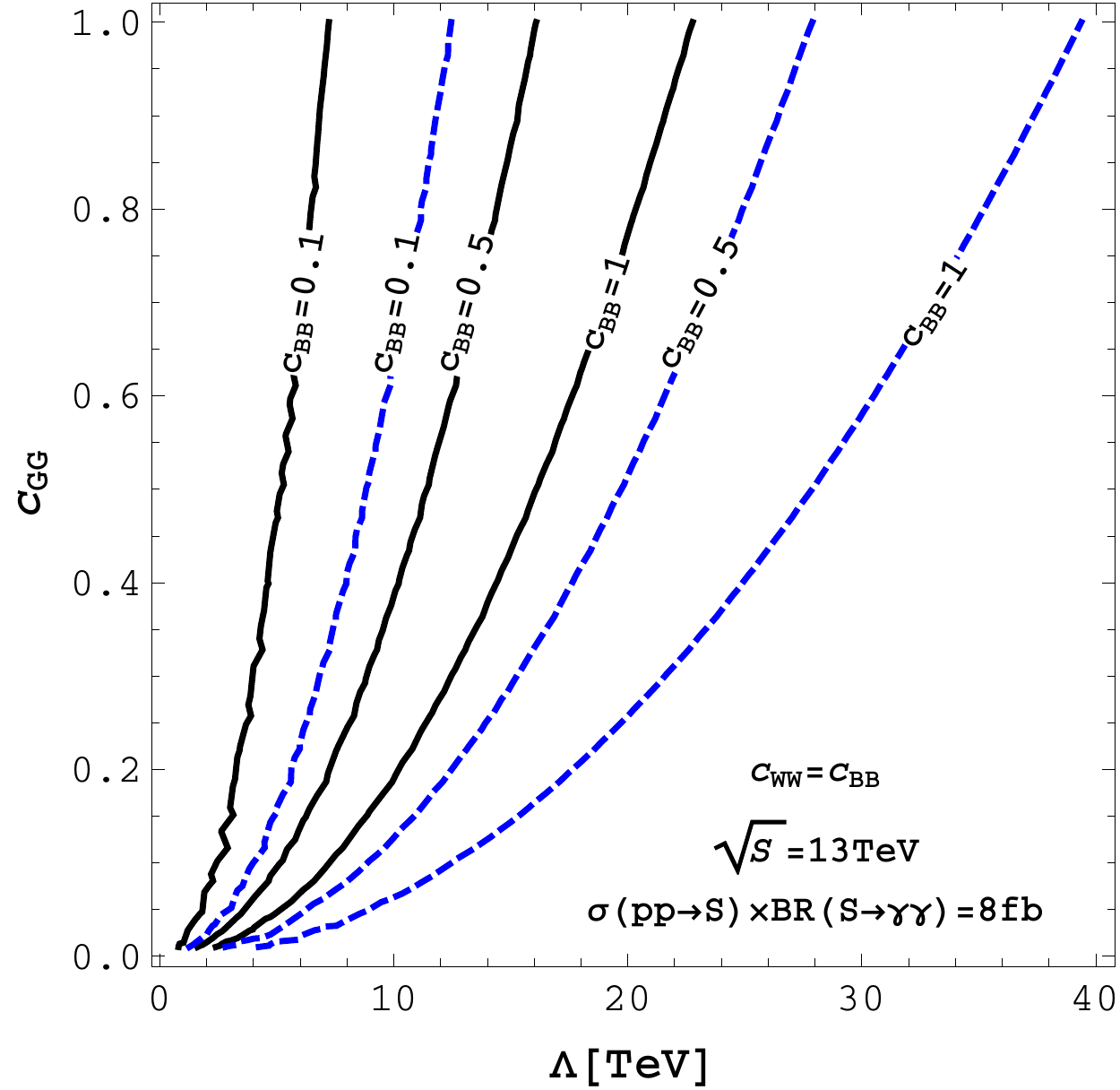}
  \caption{The allowed regions of the parameter space of our effective model from Eqs.~(\ref{xsec8})--(\ref{width8}). In the left panel, the shaded yellow area translates to the bounds of Eq.~(\ref{bounds3}). In the right panel, the curves show the compatible $(\Lambda,c_{GG})$ parameters with the diphoton cross section~\cite{Franceschini:2015kwy} fixing $c_{WW}=c_{BB}$ for three different $c_{BB}$ values: 0.1 for the leftmost curves, 0.5 for the middle curves and 1 for the rightmost ones. The solid curves represent the wide resonance case with $\Gamma_S=45$ GeV, while the dashed ones the narrow case with $\Gamma_S=5$ GeV.}
  \label{fig:param}
\end{figure}

\section{Effective couplings with SM bosons}
We assume that $S$-cion is either a CP-even or a CP-odd real scalar and parametrize its effective couplings to the SM gauge bosons in a gauge invariant way as follows 
\begin{eqnarray}
{\cal L}_{0^+} &=& \frac{c_{GG}}{\Lambda}SG^{a\mu\nu}G^a_{\mu\nu}+\frac{c_{WW}}{\Lambda}SW^{i\mu\nu}W_{i\mu\nu}+
\frac{c_{BB}}{\Lambda}SB^{\mu\nu}B_{\mu\nu}\;\; ,\;\; \hbox{CP-even}\\
{\cal L}_{0^-} &=& \frac{c_{GG}}{\Lambda}SG^{a\mu\nu}\tilde{G}^a_{\mu\nu}+\frac{c_{WW}}{\Lambda}SW^{i\mu\nu}\tilde{W}_{i\mu\nu}+
\frac{c_{BB}}{\Lambda}SB^{\mu\nu}\tilde{B}_{\mu\nu}\;\; ,\;\; \hbox{CP-odd}
\label{lagrangian}
\end{eqnarray}
where $G^{a\mu\nu}$, $W^{i\mu\nu},\; i=1,2,3$ and $B^{\mu\nu}$ are the field strength tensors of the $SU(3)$, $SU(2)_L$ and $U(1)_Y$ gauge groups, respectively, and $\tilde{G}^a_{\mu\nu}=\frac{1}{2}\varepsilon_{\mu\nu\alpha\beta}G^{a\alpha\beta}$, and similarly for $\tilde{W}$ and $\tilde{B}$. The couplings to gluons are controlled by $c_{GG}$, while $c_{BB}$ and $c_{WW}$ parametrize the couplings to the $Z$, $W$ and photons. The scale of the new physics in this EFT approach is $\Lambda$. 

The partial widths of the $S$ decays into gauge bosons are given by the following formulas
\begin{eqnarray}
\Gamma_{\gamma\gamma} &=& \frac{M_S^3}{4\pi\Lambda^2}\times (c_{BB}\cos^2\theta_W+c_{WW}\sin^2\theta_W)^2\;\; ,\;\; \Gamma_{gg} = \frac{M_S^3}{4\pi\Lambda^2}\times 8c_{GG}^2 \\
\Gamma_{ZZ} &=& \frac{M_S^3}{4\pi\Lambda^2}\beta_Z\delta_Z\times (c_{WW}\cos^2\theta_W+c_{BB}\sin^2\theta_W)^2\;\; ,\;\;\Gamma_{WW} = \frac{M_S^3}{4\pi\Lambda^2}\beta_W\delta_W\times 2c_{WW}^2 \\
\Gamma_{Z\gamma} &=& \frac{M_S^3}{4\pi\Lambda^2}(1-r_z)^3\times 2\cos^2\theta_W\sin^2\theta_W(c_{BB}-c_{WW})^2
\label{widths}
\end{eqnarray}
where $M_S=750$ GeV is the mass of $S$-cion, $r_{W,Z}=m_{W,Z}^2/M_S^2$, $m_Z=91.18$ GeV, $m_W=80.39$ GeV, $\beta_{W,Z}=\sqrt{1-4r_{W,Z}}$ and $\delta_{W,Z}=1-4r_{W,Z}+6r_{W,Z}^2\eta$; $\sin^2\theta_W=0.233$ and $\cos^2\theta_W=1-\sin^2\theta_W$. CP-even an CP-odd scalars have identical partial widths to $\gamma\gamma$, $Z\gamma$ and $gg$, but not to $ZZ,WW$. For CP-even $S$-cions $\eta=+1$, and for CP-odd, $\eta=0$ in the above $\delta_{W,Z}$ formulas. However, as $r_{W,Z}\sim 10^{-2}$, the difference between CP-even and odd for $ZZ$ and $WW$ is negligible. From now on we use $S$ to denote either a CP-even or a CP-odd $S$-cion unless explicitly stated otherwise.

Assuming these interactions, the diphoton signal is the production of $S$ in gluon fusion with subsequent decay to a pair of photons. The ATLAS data shows an $\sim 3.6\sigma$ excess with 14 events observed while the CMS data an  $\sim 2.6\sigma$ excess with 10 events. We will assume $\sigma(pp\to S)\times BR(S\to\gamma\gamma)=8$ fb~\cite{atlasresult, cmsresult} as the signal cross section in the rest of the work. Nevertheless, in Refs.~\cite{Falkowski:2015swt,Buckley:2016mbr} we find that fittings to the Run I and Run II data might actually accommodate narrower resonances and smaller cross sections. We thus investigate a range of total widths, from narrow ($\Gamma_S=5$ GeV) to wide ($\Gamma_S=45$ GeV). We will see that the results depend sensibly to the size of the total width.

\subsection{Constraints on $c_{GG}$, $c_{WW}$ and $c_{BB}$}

In this subsection, we discuss the constraints on the coefficients appearing in the EFT. The first constraint comes from fitting the observed signal cross section of $\sigma(pp\to S)\times BR(S\to\gamma\gamma)$. In order to adjust the gluon fusion production cross section we need to pick a consistent value of $c_{GG}$. The $\Lambda$ and $c_{GG}$ parameters (as well as $c_{BB}$ and $c_{WW}$) are constrained by the observed signal of the diphoton channel and by the estimate of the total width of the $S$-cion
\begin{eqnarray}
&& \sigma(pp\to S)\times BR(S\to\gamma\gamma)=8\; \hbox{fb}\label{xsec8}\\
&& 5\; \hbox{GeV} \leq \sum_{k} \Gamma_k  \leq 45\; \hbox{GeV}
\label{width8}
\end{eqnarray} 
where $\Gamma_k$ represent all the partial widths of the $S$ decay to gauge bosons and possibly other states about whose identity we remain agnostic.

In Run I, the ATLAS and CMS Collaborations searched for a Higgs boson decaying to $ZZ$~ \cite{ATLAS:2015hb}, $W^+W^-$~ \cite{CMS:2015vv, ATLAS:2015sf}, $Z\gamma$~\cite{ATLAS:1407.8150} and $\gamma\gamma$~\cite{CMS-PAS-HIG-14-006, Aad:2015mna} with no significant excesses observed in any channel. These searches constrain the $c_{WW}$ and $c_{BB}$ couplings. We now describe these bounds in some detail.

In \cite{ATLAS:2015hb}, the search for a scalar $S$ produced by gluon fusion and decaying to $ZZ$ was conducted in several final state channels: $\ell \ell \ell \ell$, $\ell \ell \nu \nu$, $\ell \ell q q$, and $\nu \nu q q$. The combined limit for $m_S = 750$ GeV was obtained as
\begin{equation}
\sigma (gg \rightarrow S)_{\sqrt{s}=8 \,\, {\rm TeV}} \times {\rm Br}(S \rightarrow ZZ) \, < \, 12 \,\,{\rm fb} \,\,.
\end{equation}
This limit can be translated into a bound at $\sqrt{s} = 13$ TeV by scaling the production cross section to its appropriate value. This is done by scaling the dimensionless partonic integral $C_{gg}$ at $m_S = 750$ GeV from its value at 8 TeV to its value at 13 TeV. One has, using MSTW2008NLO pdfs evaluated at the energy scale of 750 GeV, that $C_{gg, 8 \,\, TeV} = 174$ while $C_{gg, 13 \,\, TeV} = 2137$. Then, the gain factor is $[C_{gg}/s]_{13 \,\, TeV}/ [C_{gg}/s]_{8 \,\, TeV} \,= \, 4.7$. This implies that 
\begin{equation}
\sigma (gg \rightarrow S)_{\sqrt{s}=13 \,\, {\rm TeV}} \times {\rm Br}(S \rightarrow ZZ) \, < \, 56.4 \,\,{\rm fb} \,\,.
\end{equation}
Comparing this to the signal in Eq.~\ref{xsec8}, we obtain 
\begin{equation}
\frac{\Gamma_{ZZ}}{\Gamma_{\gamma\gamma}} < 7 \,\,.
\end{equation}

Similarly, \cite{ATLAS:2015sf} constrained a scalar $S$ produced by gluon fusion and decaying to $WW$ in the final state of  $\ell \nu q q$ at 8 TeV, obtaining
\begin{equation}
\sigma (gg \rightarrow S)_{\sqrt{s}=8 \,\, {\rm TeV}} \times {\rm Br}(S \rightarrow WW) \, < \, 40 \,\,{\rm fb} \,\,.
\end{equation}
Using the gain factor of $[C_{gg}/s]_{13 \,\, TeV}/ [C_{gg}/s]_{8 \,\, TeV} \,= \, 4.7$, the bound can be translated to
\begin{equation}
\sigma (gg \rightarrow S)_{\sqrt{s}=13 \,\, {\rm TeV}} \times {\rm Br}(S \rightarrow WW) \, < \, 188 \,\,{\rm fb} \,\,.
\end{equation}
Comparing to Eq.~\ref{xsec8}, one obtains
\begin{equation}
\frac{\Gamma_{WW}}{\Gamma_{\gamma\gamma}} < 23 \,\,.
\end{equation}

The most stringent bounds on the coefficients come from scalars decaying to $Z \gamma$. From \cite{ATLAS:1407.8150}, one obtains 
\begin{equation}
\sigma (gg \rightarrow S)_{\sqrt{s}=8 \,\, {\rm TeV}} \times {\rm Br}(S \rightarrow Z \gamma ) \, < \, 3.4 \,\,{\rm fb} \,\,.
\end{equation}
Using the gain factor and the signal rate from Eq.~\ref{xsec8}, one obtains
\begin{equation}
\frac{\Gamma_{Z\gamma}}{\Gamma_{\gamma\gamma}} < 2 \,\,.
\end{equation}

Summarizing the above, we thus have
%
%
\begin{equation}
\frac{\Gamma_{ZZ}}{\Gamma_{\gamma\gamma}} < 7,\;\; \frac{\Gamma_{WW}}{\Gamma_{\gamma\gamma}} < 23,\;\; 
\frac{\Gamma_{Z\gamma}}{\Gamma_{\gamma\gamma}} < 2 \,\,.
\label{bounds1}
\end{equation}

\begin{figure}[b]
  \centering
  \includegraphics[width=.2\textwidth]{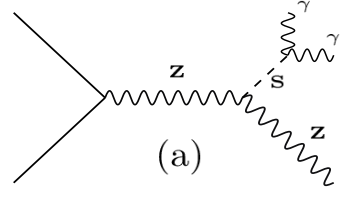}
  \includegraphics[width=.2\textwidth]{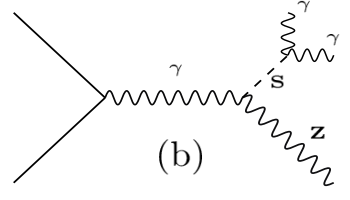}
  \includegraphics[width=.2\textwidth]{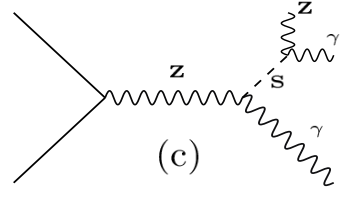}
  \includegraphics[width=.2\textwidth]{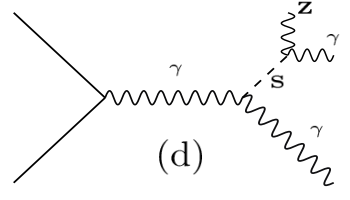}\\
  \includegraphics[width=.2\textwidth]{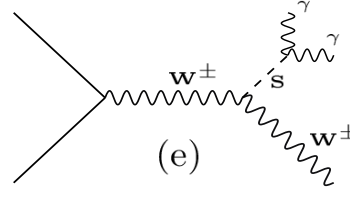}
  \includegraphics[width=.2\textwidth]{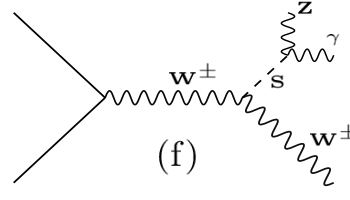}
  \caption{The diagrams $(a)$--$(d)$ contribute to the signal process $q\bar{q}\to \gamma\gamma\ell^+\ell^-$. The diagrams $(a)$ and $(b)$ involve the scattering amplitudes $(S\to\gamma\gamma)+Z$, while $(c)$ and $(d)$ the amplitudes for $(S\to Z\gamma)+\gamma$. The $Z$ bosons decay to electrons and muons. The diagram $(e)$ represents the amplitude for $q\bar{q}\to S+W^\pm\to \gamma\gamma\ell^\pm+\eslash$, and $(f)$ the process $q\bar{q}\to S+W^\pm\to \ell^\pm\ell^\mp\ell^\pm+\gamma+\eslash$. The $W$ bosons also decay leptonically. Here $S$ denotes either a scalar or pseudoscalar.}
  \label{fig:feyn}
\end{figure}

In the left panel of Figure~(\ref{fig:param}) we display the allowed region of the $c_{WW}\times c_{BB}$ plane considering the most stringent constraint from Eq.~(\ref{bounds1}): $\Gamma_{Z\gamma}/\Gamma_{\gamma\gamma}<2$. This bound translates effectively to the constraint
\begin{equation}
-0.6 < \frac{c_{WW}}{c_{BB}} < 6.4
\label{bounds3}
\end{equation} 
In the right panel of Figure~(\ref{fig:param}) we show the $(\Lambda, c_{GG})$ combinations compatible with the observed signal in the diphoton channel of Eq.~(\ref{xsec8},\ref{width8})~\cite{Franceschini:2015kwy}. We see that we can safely lie within the perturbative regime of $c_{GG}/\Lambda$ and the validity region of the EFT where $\Lambda > M_S$. Constraints on new scalar and pseudoscalar particles in different contexts can be found in Refs.~\cite{Carmi:2012yp,Carmi:2012in,Jaeckel:2012yz}.

\section{Results}

We simulated parton-level events with \texttt{MadGraph5}~\cite{mad5} for the new physics signals 
\begin{eqnarray}
pp &\to& S+Z\to \gamma\gamma+\ell^+\ell^-\label{aall}\\
pp &\to& S+W^\pm\to\gamma\gamma+\ell^\pm+\eslash\label{aalet}\\
pp &\to& S+W^\pm\to \ell^\pm\ell^\mp\ell^\pm+\gamma+\eslash\label{alllet}
\end{eqnarray}
where $\ell$ denotes electrons end muons, and SM backgrounds at the 13 TeV LHC. 

For the signals we chose the following simulation parameters: $\Lambda=1$ TeV for all processes,  $c_{WW}=c_{BB}=c_{GG}=1$ in the generation of the events of the process (\ref{aalet}), and $\frac{1}{2}c_{WW}=c_{BB}=c_{GG}=1$ in the generation of the events of the process (\ref{alllet}). For the signals of type of the Eq.~(\ref{aall}), contrary to the other ones, a simple rescaling to compute cross sections for different parameters does not work. In this case we ran events for several points of the parameters space in order to obtain our results. We point out that these parameters were chosen only to generate events for the purpose to estimate the cut efficiencies. The cross sections where subsequently rescaled for realistic parameters represented in our results.

The signal cross section for these simulated models and the SM backgrounds are displayed in Table I. We do not take detector effects into account, but as a cut-and-count analysis within a very clean environment with very hard photons and leptons, we expect that our estimates will not change too much in a more careful simulation. Yet, simulating detectors, extra jets radiation, and hadronization will be important to check these parton-level results.

We show in Figure~(\ref{fig:feyn}) the Feynman diagrams for the processes $pp \to S+Z\to \gamma\gamma+\ell^+\ell^-$, represented by the amplitudes $(a)$--$(d)$, $pp \to S+W^\pm\to\gamma\gamma+\ell^\pm+\eslash$, represented by the amplitude $(e)$, and $pp \to S+W^\pm\to \ell^\pm\ell^\mp\ell^\pm+\gamma+\eslash$ corresponding to diagram $(f)$. The signals involve hard photons and/or leptons and missing energy if the process contains a $W$ boson. The diagrams $(c)$ and $(d)$ contribute little to $\gamma\gamma\ell^+\ell^-$ due the hard $M_{\gamma\gamma}$ cut. We point out that the $S$-cion decays to diphotons and $Z\gamma$ to produce the $\gamma\gamma\ell^+\ell^-$  final states.

The relevant SM backgrounds for $pp\to SZ$ are $Z\gamma\gamma$, $Zj\gamma$ and $Zjj$, and for $pp\to S+W^\pm$, by its turn, are $W^\pm\gamma\gamma$, $W^\pm j\gamma$, $W^\pm jj$ and $Z\gamma\gamma$. The three lepton signal from $pp\to S+W^\pm S+W$ has as the dominant backgrounds $ZW^\pm\gamma$ and $ZW^\pm j$. For background events with jets we assumed $P_{j\to\gamma}=1.2\times 10^{-4}$ as the jet mistagging probability. The photon and lepton ID efficiencies were taken as $\varepsilon_\gamma=0.95$ and $\varepsilon_\ell=0.95$, respectively. We multiplied all the backgrounds by their QCD K-factors~\cite{mad5}, see Table~\ref{tab}.
\begin{table}
\begin{center}
\begin{tabular}{l|l|l|l}
\hline
Process & $\sigma$(fb) & K-factor & $\varepsilon_{cut}$ \\
\hline
Signal $\gamma\gamma\ell^+\ell^-$ & 1.47(1.43) & 1 & 0.76 \\	
$Z\gamma\gamma$ &  0.39 & 1.51 & 0.01 \\
$Z j\gamma$ & 148.0 & 1.47 & 0.01 \\	
$Zjj$ & $37.34\times 10^3$ & 1.14 & 0.03 \\
\hline
Signal $\gamma\gamma\ell^\pm\eslash$ & 18.6(14.9) & 1 & 0.88 \\
$W^\pm\gamma\gamma$ &  0.39 & 3.1 & 0.10 \\
$W^\pm j\gamma$ & 948.5 & 1.45 & 0.02 \\	
$W^\pm jj$ & $228\times 10^3$ & 1.15 & 0.08 \\
\hline
Signal $\ell\ell\ell\gamma\eslash$ & 0.91(0.83) & 1 & 0.96 \\
$ZW^\pm\gamma$ &  0.13 & 2.34 & 0.10 \\
$ZW^\pm j$ & 49 & 1.30 & 0.13 \\	
\hline
\end{tabular}
\caption{The cross sections (in fb) after imposing the basic cuts of Eq.~(\ref{bcuts}), the QCD K-factors used in the signal and backgrounds normalizations, and the cut efficiencies for signals and backgrounds (in the rightmost column) after imposing the harder cuts of Eq.~(\ref{hcuts}) and (\ref{3lcut}). The signal cross sections and cut efficiencies correspond to the benchmark point. The numbers outside(inside) parenthesis in the signal rows correspond to the CP-even(odd) cross sections. The signal efficiencies are the same independently the type of $S$-cion. The cross section for $\gamma\gamma\ell^+\ell^-$ in the first row assumes $c_{WW}=c_{BB}=1$.}
\label{tab}
\end{center}
\end{table}

In Table I we also display the cut efficiencies for signals and backgrounds after demanding the following basic cuts for the $\gamma\gamma\ell\ell$ events
\begin{eqnarray}
&& p_{T_\gamma}>30\; \hbox{GeV},\;\; p_{T_\ell}>30\; \hbox{GeV},\;\; |\eta_{\gamma,\ell}|<2.5\nonumber\\
&& \Delta R_{\gamma\gamma}>0.4,\;\; \Delta R_{\gamma\ell}>0.4,\;\; \Delta R_{\ell\ell}>0.4\nonumber\\
&& \eslash\; > 20\; \hbox{GeV}\;\;\;\; \hbox{for events with neutrinos}
\label{bcuts}
\end{eqnarray}
The $\eslash$ cut is aimed to trigger on missing energy events. 

We further impose hard cuts to separate signals from backgrounds taking advantage of the hard $\gamma\gamma$ spectrum of the $S$-cion decays and that charged leptons pairs are produced through the $Z$ boson decay. 
\begin{equation}
p_{T_\gamma}>100\; \hbox{GeV},\;\; |M_{\ell\ell}-m_Z|<20\; \hbox{GeV},\;\; M_{\gamma\gamma} > 500\; \hbox{GeV}
\label{hcuts}
\end{equation} 
The cut on the invariant mass of $M_{\ell\ell}$  is not imposed to events containing $W$ bosons. Multi-photon and SM Higgs backgrounds are negligible after cuts.

For the three leptons signal $\ell\ell\ell\gamma\eslash$, instead of imposing a cut on the photons invariant mass $M_{\gamma\gamma}$, we demand
\begin{equation}
M_{\gamma\ell\ell} > 500\;\hbox{GeV}
\label{3lcut}
\end{equation}

\begin{figure}[!t]
  \centering
  \includegraphics[width=.37\textwidth]{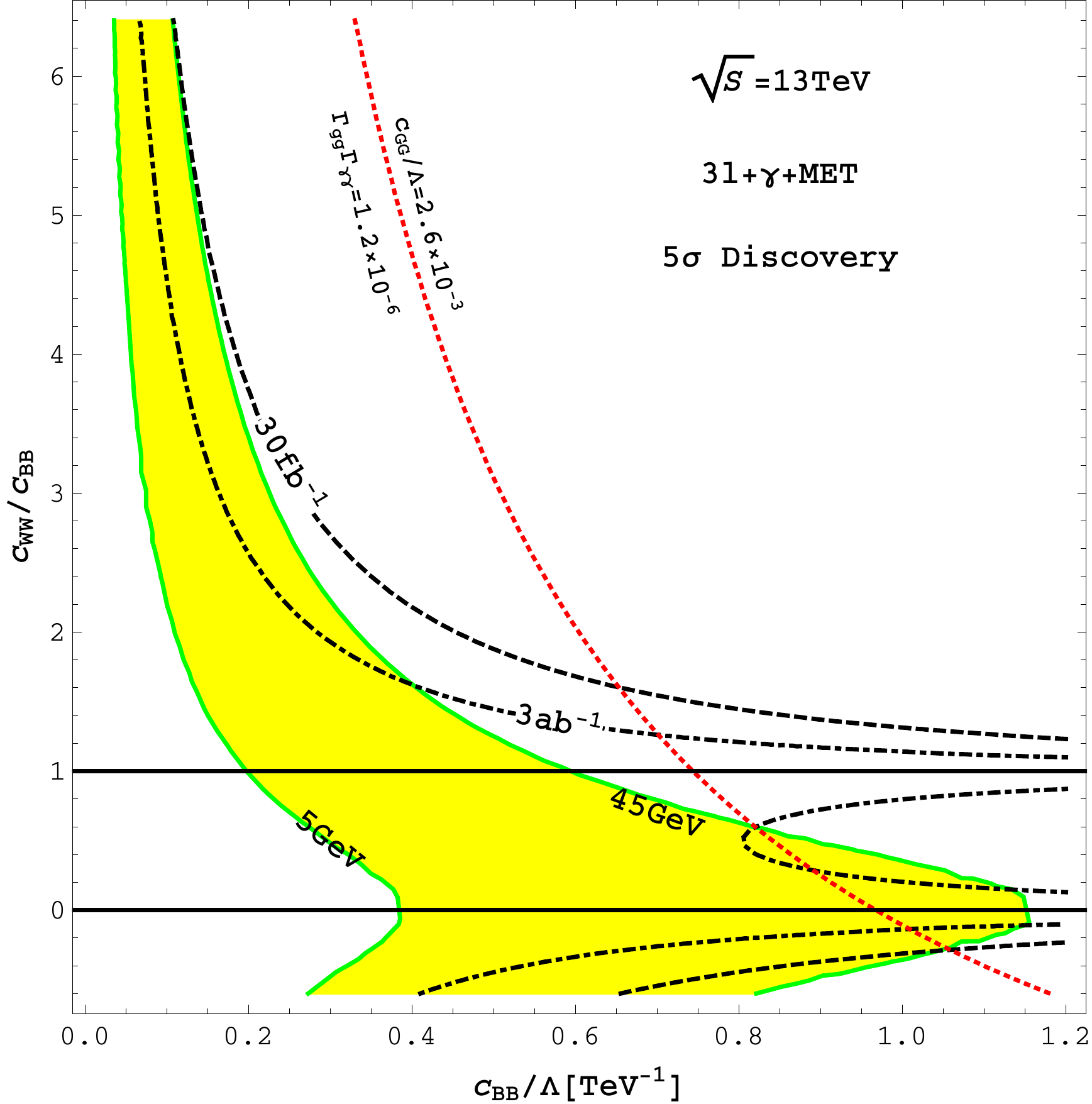}
  \includegraphics[width=.37\textwidth]{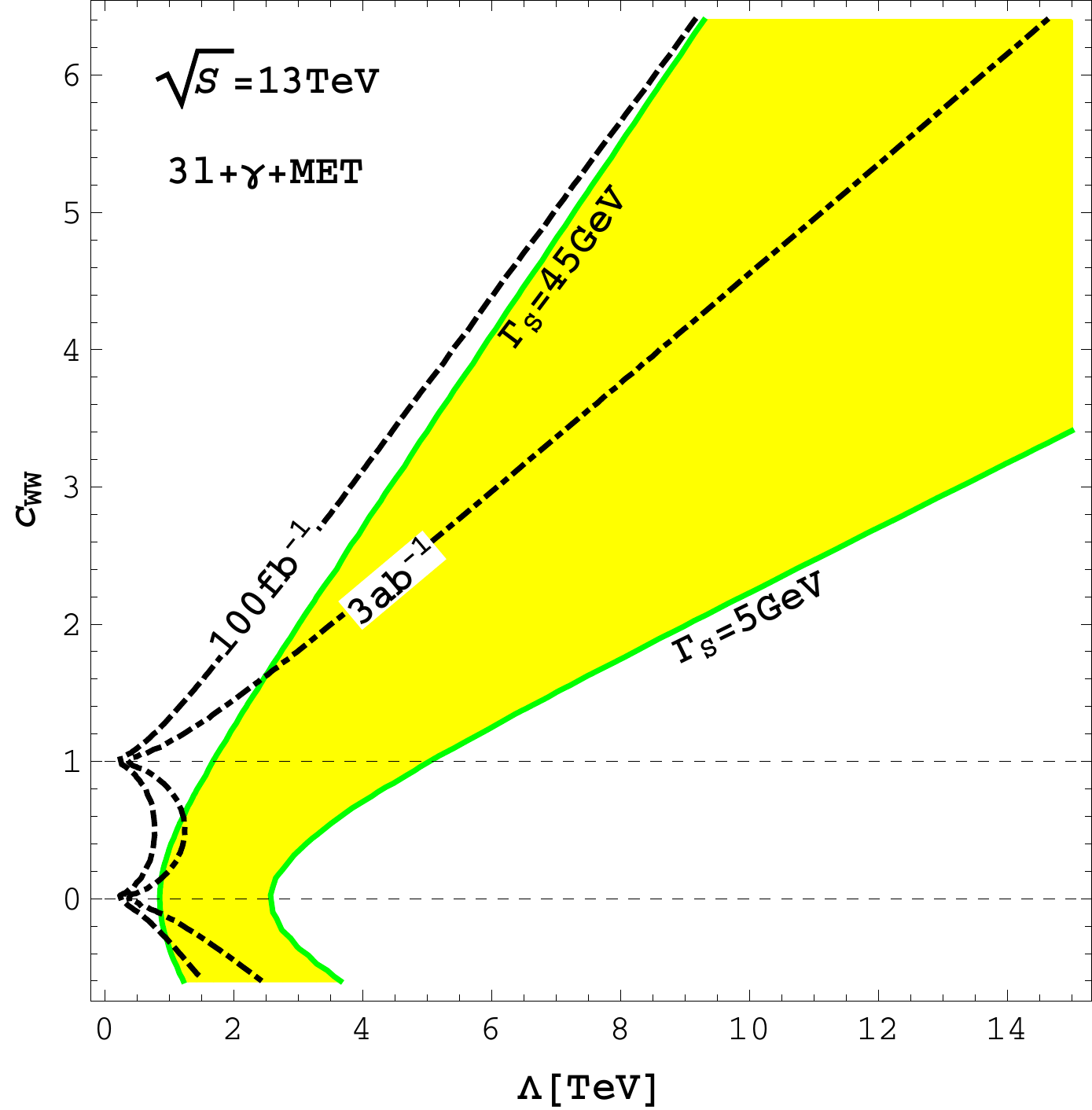}\\
  \includegraphics[width=.37\textwidth]{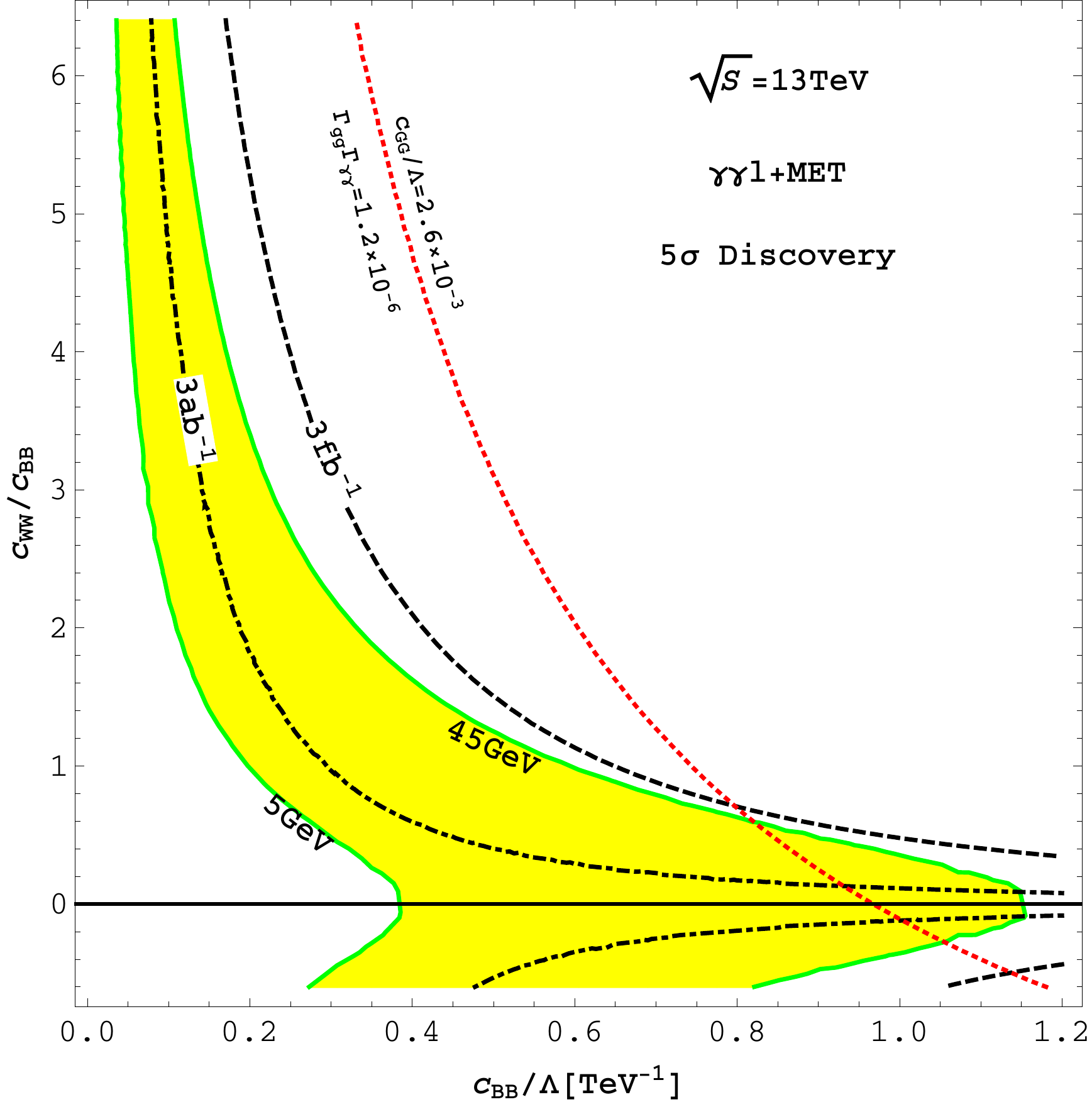}
  \includegraphics[width=.37\textwidth]{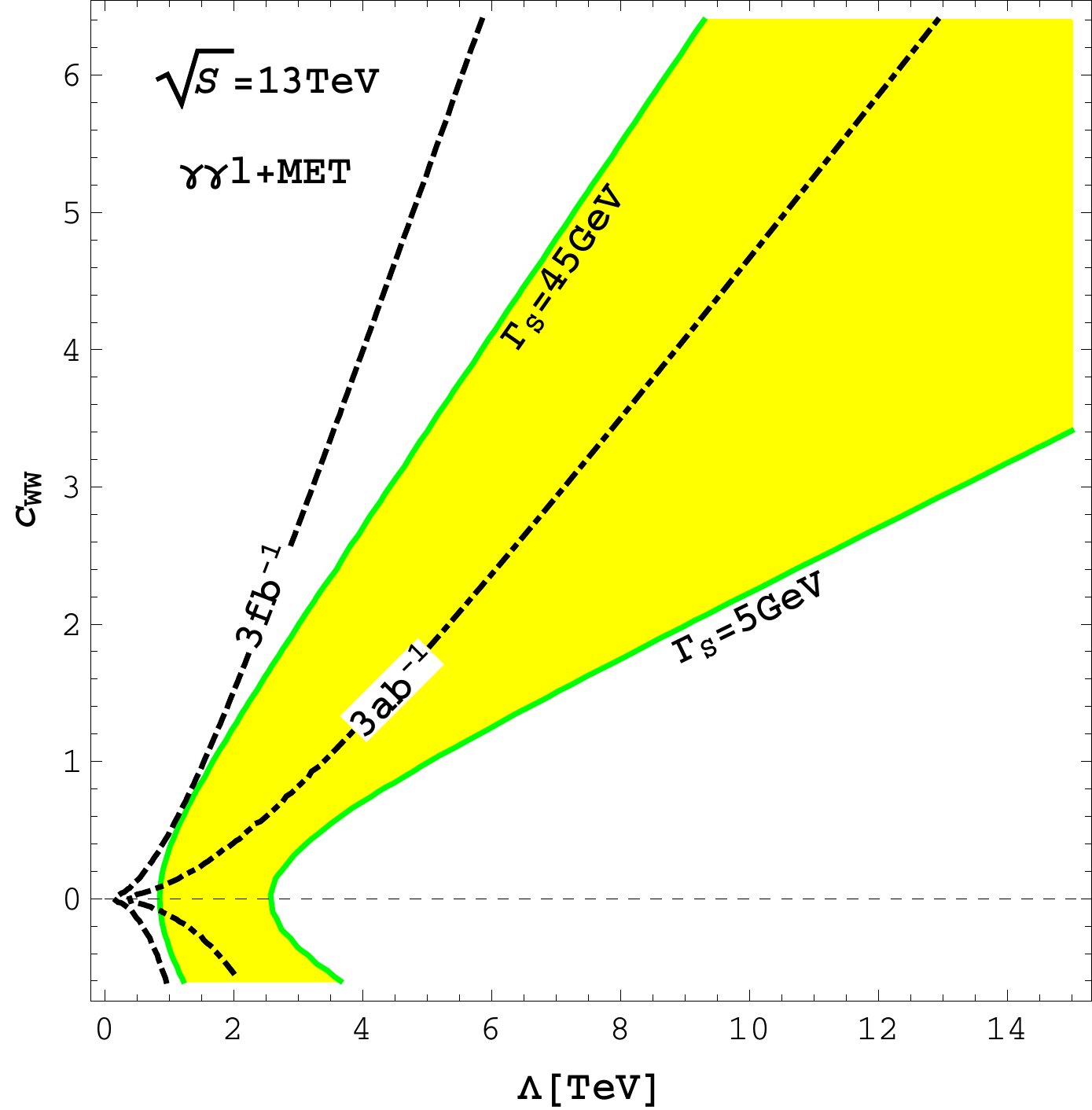}\\
  \includegraphics[width=.37\textwidth]{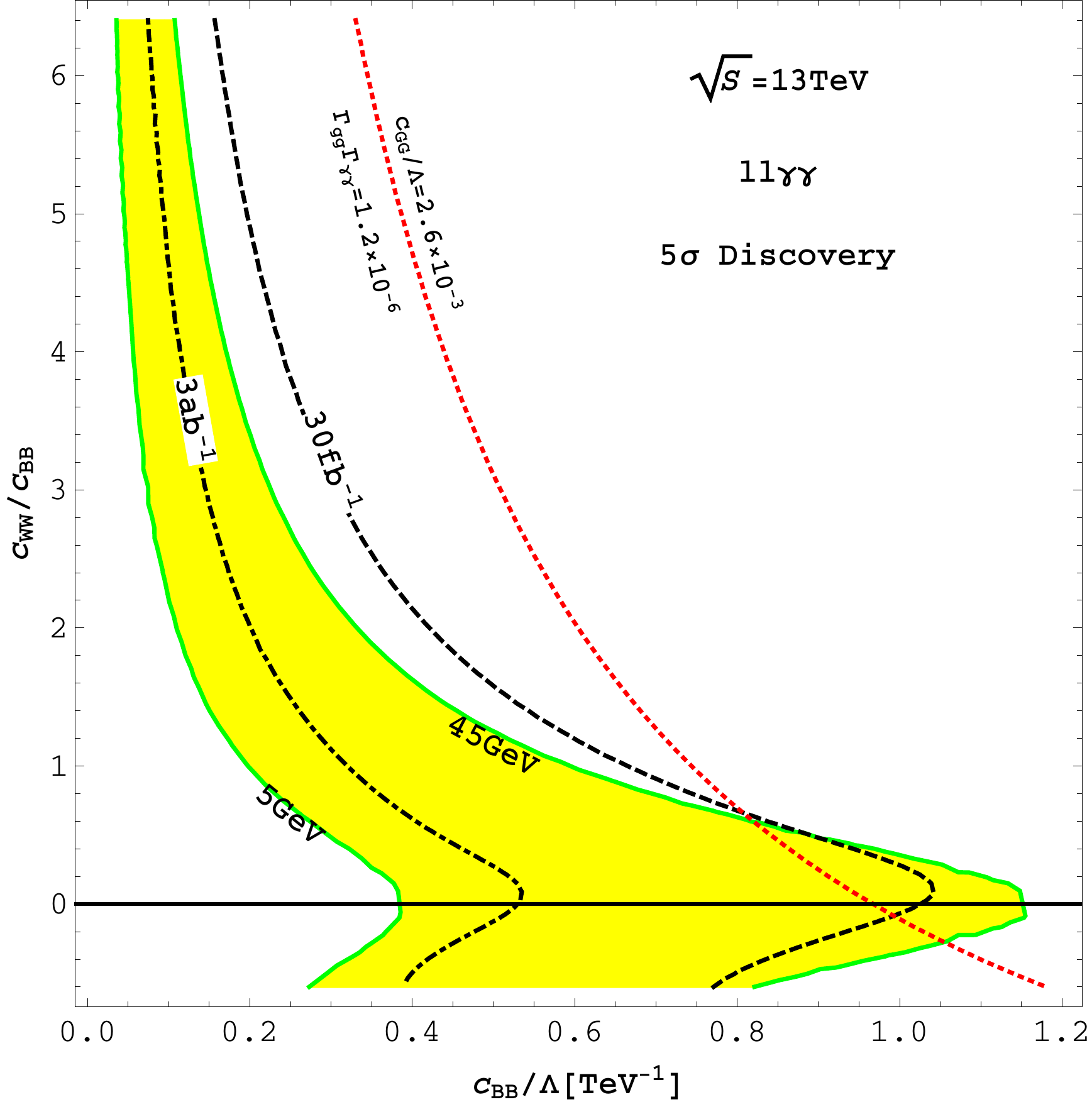}
  \includegraphics[width=.37\textwidth]{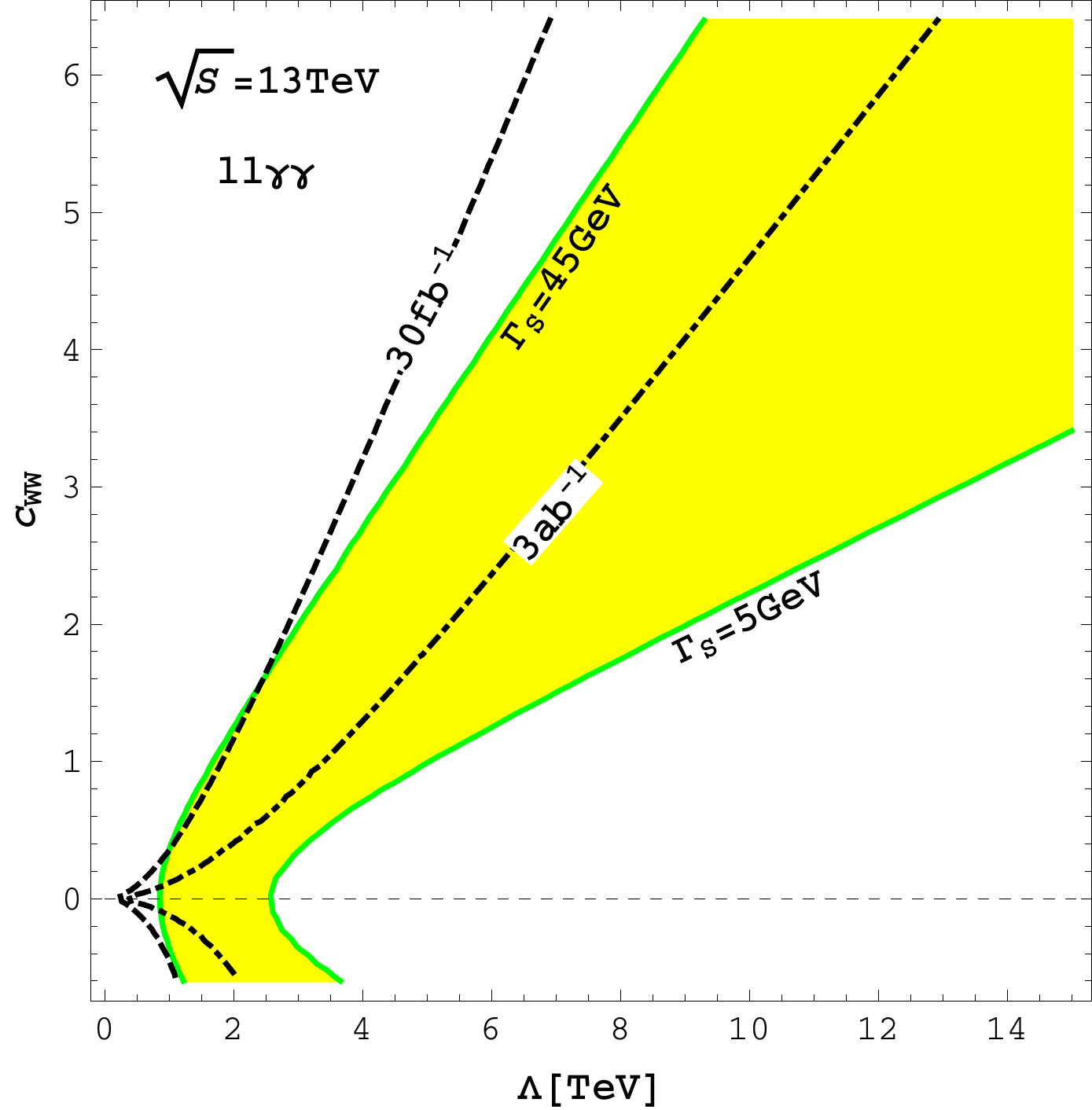}\\
  \caption{Discovery reach of the LHC 13 TeV for 3 up to 3000 fb$^{-1}$ in the $c_{WW}/c_{BB}$ {\it versus} $c_{BB}/\Lambda$ plane (left panels) in the $\gamma\gamma\ell\ell$ (lower), $\gamma\gamma\ell\eslash$ (middle) and $\ell\ell\ell\gamma\eslash$ (upper) channels. The right panels show the same results in the $c_{WW}$ {\it versus} $\Lambda$ space with $c_{BB}=1$ fixed. 
Results for the CP-odd case are essentially the same. The yellow shaded area between solid green lines represent the points where the total width varies from 5 to 45 GeV. The dotted red line are the points respecting $\sigma(pp\to S\to\gamma\gamma)=8$ fb with $c_{GG}/\Lambda=2.6\times 10^{-3}$ fixed as an example. Solutions lie on the dotted red line and inside the shaded area and can be discovered at 5$\sigma$ if they also lie between the black dashed and black dot-dashed lines with luminosities from 3 up to 3000 fb$^{-1}$.}
  \label{figzw}
\end{figure}

The $5\sigma$ discovery regions, respecting all the constraints, for various integrated luminosities from $L=$3 up to 3000 fb$^{-1}$ at the 13 TeV LHC are shown in the upper, middle and lower panels of Figure~(\ref{figzw}) for the $\ell\ell\ell\gamma\eslash$, $\gamma\gamma\ell\eslash$ and $\gamma\gamma\ell\ell$ channels, respectively, in the $c_{WW}/c_{BB}$ {\it versus} $c_{BB}/\Lambda$ space. We see that there is already room for a new signal discovery with the current data in the $\gamma\gamma\ell\ell$ channel if the scalar width is around 45 GeV, as suggested by the ATLAS data.  The results displayed in Figure~(\ref{figzw}), and in the subsequent figures actually, are nearly the same for CP-even and CP-odd scalar once the CP-odd scalars have around 10\% smaller cross sections compared to the CP-even case. 
%

 In the left panels of Figure~(\ref{figzw}), the dotted red lines represent points satisfying Eq.~(\ref{xsec8}). The yellow region denotes the parameter space that respects Eq.~(\ref{width8}) and is bounded by solid green lines to the right(left) which represent the wide(narrow) resonance cases of 45(5) GeV. The points of this parameter space compatible with the diphoton data are thus those on the red dotted line and inside the region delimited by the green lines. We fixed $c_{GG}/\Lambda=2.6\times 10^{-3}$ in the red lines, but we checked that it is possible to find simultaneous solutions to Eqs.(\ref{xsec8}) and (\ref{width8}) for all points inside the yellow shaded area by adjusting the relevant parameters respecting all collider bounds with $c_{GG}/\Lambda$ from $2.2\times 10^{-3}$ to $8\times 10^{-3}$. The regions between the dashed and dot-dashed black lines are those than can be discovered at $5\sigma$ significance by the LHC up to 3 ab$^{-1}$. For example, in the channel $\ell\ell\ell\gamma\eslash$, shown in the upper plots, the dotted red line crosses the 45 GeV line at two points: one for $c_{WW}/c_{BB}\approx 0.6$ where around 3 ab$^{-1}$ will be necessary for a $5\sigma$ discovery, the other where $c_{WW}/c_{BB}\approx -0.2$, which can be probed with much less luminosity, around 100 fb$^{-1}$.

For the $\ell\ell\ell\gamma\eslash$ signal, we see that points close to $c_{WW}/c_{BB}=0$ or 1 cannot be accessed by the LHC. This can be easily understood by noting that the cross section for this channel vanishes if $c_{WW}=0$ or $c_{WW}=c_{BB}$. However, these gaps can be probed in the other two proposed channels. From the middle left panel, we actually see that $\gamma\gamma\ell\eslash$ events can already be found in the current data for $c_{WW}\approx c_{BB}$ if the resonance is wide, but will give no signals if $c_{WW}=0$ once the $SWW$ coupling is involved in its production mechanism. The $\gamma\gamma\ell\ell$ channel, in turn, can be observed for the entire allowed $c_{WW}/c_{BB}$ region with luminosities up to 3 ab$^{-1}$, again, if the resonance is wide enough. The three channels are thus complementary and should be looked for in an experimental study.

A general feature of these search channels is that the width of the resonance should be wide enough. For $\Gamma_S\sim 5$ GeV the LHC cannot observe events from these channels. A wide resonance composed of SM decays only implies larger couplings of the $S$-cion with the gauge bosons which would enable the LHC to probe the channels studied in this work.

In the right panels of Figure~(\ref{figzw}) we show the same results in the $c_{WW}$ {\it versus} $\Lambda$ plane but fixing $c_{BB}=1$. For $c_{WW}$ up to the saturating value of Eq.~(\ref{bounds3}) a new physics scale up to 13 TeV can be probed in these channels with 3 ab$^{-1}$ for a resonance of width not smaller than $\sim 20$ GeV. For the wide resonance scenario of 45 GeV, scales up to 9 TeV can be probed at the $5\sigma$ statistical level in the $\gamma\gamma\ell\ell$($\gamma\gamma\ell\eslash$)[$\ell\ell\ell\gamma\eslash$] process with 100(280)[65] fb$^{-1}$ and a $3\sigma$ evidence can be reached with much less luminosity of 40(100)[25] fb$^{-1}$. Spectacularly, in the process $\ell\ell\ell\gamma\eslash$, $\Lambda=1.5$(3) TeV is accessible with the present available data for discovery(evidence). In Table~(\ref{restab}) we summarize the reach of the LHC to probe new physics related to the $S$-cion for $c_{WW}$ up to the saturating bound of Eq.~(\ref{xsec8}--\ref{width8}) for various luminosities. We also quote in the first row of Table~(\ref{restab}) the $\Lambda$ scales which would produce a 3$\sigma$ evidence for a $S$-cion with the present data.

%
%

\begin{table}
\begin{center}
\begin{tabular}{c|c|c|c}
\hline
Channel & Luminosity (fb$^{-1}$) & $\Lambda$(TeV) & $\Gamma_S$(GeV) \\
\hline
$\gamma\gamma\ell\ell$   & 100(40) & 9 & 45 \\
                         & 3000 & 14 & 18 \\
\hline	
$\gamma\gamma\ell\eslash$ & 280(100) & 9.4 & 45 \\
                        & 3000 & 13.3 & 22 \\
\hline
$\ell\ell\ell\gamma\eslash$ & 3 & 1.5(3) & 45 \\
                         & 65(25) & 9.3 & 45 \\
                         & 3000 & 13 & 22 \\
\hline
\end{tabular}
\caption{The $\Lambda$ scale (in TeV) and minimum total width (in GeV) at which a CP-even $S$-cion can be discovered (5$\sigma$) in the 13 TeV LHC for various integrated luminosities for the three channels studied in this work. The numbers in the parenthesis represent the scales at which a 3$\sigma$ evidence would possible. CP-odd scalars present essentially the same results.}
\label{restab}
\end{center}
\end{table}

\begin{figure}[!t]
  \centering
  \includegraphics[width=1.\textwidth]{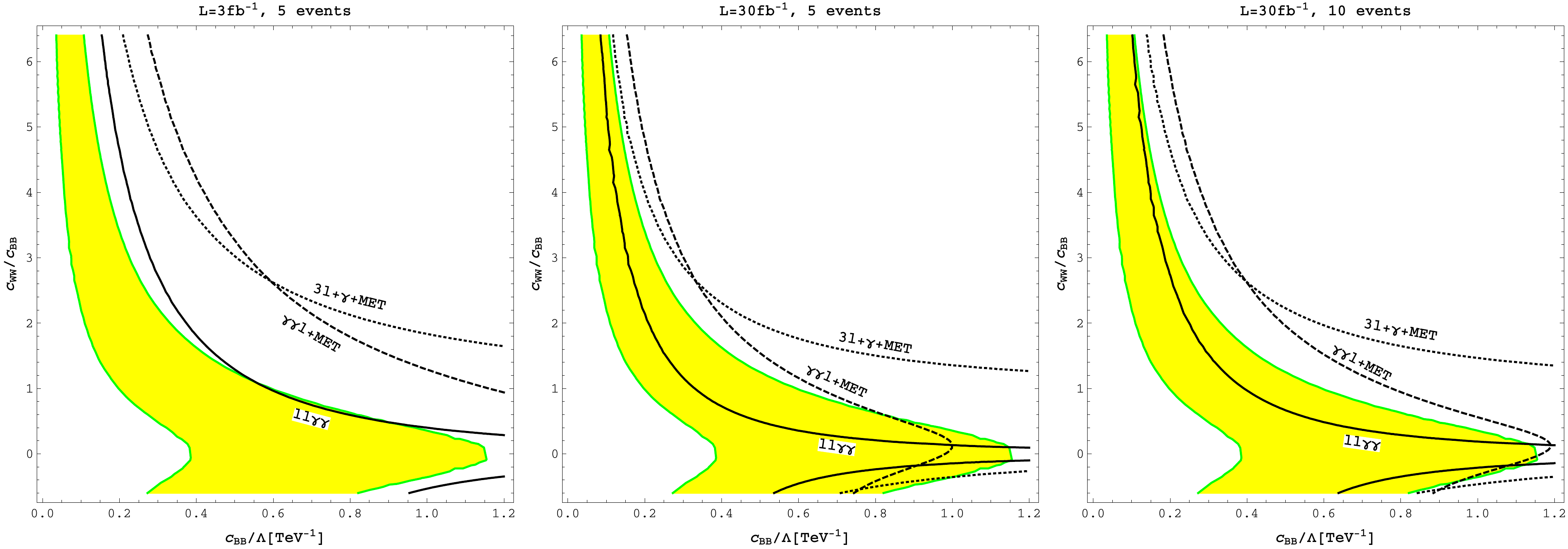}
  \caption{The lines where 5 (left and middle panels) and 10 signal events (right panel) are expected at the 13 TeV LHC after 3 fb$^{-1}$ (left) and 30 fb$^{-1}$ (middle and right) in the $\gamma\gamma\ell\ell$ (solid), $\gamma\gamma\ell\eslash$ (dashed) and $\ell\ell\ell\gamma\eslash$ (dotted) channels in the $c_{WW}/c_{BB}$ {\it versus} $c_{BB}/\Lambda$ space. The cuts of Eqs.~(\ref{bcuts}--\ref{3lcut}) and photon and lepton efficiency factors were applied in these results. The shaded yellow area between the green solid lines has the same interpretation of the ones encountered in Figure~(\ref{figzw}). Results for the CP-odd case are essentially the same.}
  \label{fignev}
\end{figure}

 For 3 fb$^{-1}$, 0.02, 0.4 and 0.07 background events are expected in the $\gamma\gamma\ell\ell$, $\gamma\gamma\ell\eslash\;$ and $\ell\ell\ell\gamma\eslash$ channels, respectively, with the cuts of Eqs.~(\ref{bcuts})--(\ref{3lcut}), and at least 5 signal $\gamma\gamma\ell\ell$ events if $0.6<c_{WW}/c_{BB}<1.2$ and $\Lambda\sim 1.5$ TeV. In Figure~(\ref{fignev}) we show our estimate of the number of signal events expected for 3 and 30 fb$^{-1}$ in the $c_{WW}/c_{BB}$ {\it versus} $c_{BB}/\Lambda$ space. 

With 30 fb$^{-1}$, 5 $\gamma\gamma\ell\ell$ events are expected for the entire region allowed by current constraints, while the same number of $\gamma\gamma\ell\eslash\;$ and $\ell\ell\ell\gamma\eslash$ events can be observed in the regions shown in Figure~(\ref{fignev}) where we also show the regions with 10 events expected. Again, the CP-odd scalars have very similar results. It would be very interesting to check whether any excess might be lurking in the Run II data in these channels.

\section{Examples of UV Completions}
It is important to have an idea of which models could be searched for in these channels. As an example, let us assume that a CP-even $S$-cion is produced through a loop of $N_Q$ vector-like quarks of charge $q_Q$ pertaining to possibly higher $SU(3)_C$ representation as suggested in Ref.~\cite{Sierra:2015zma}. Additional contributions are expected in the decay to photons from $N_L$ vector-like leptons with charges $q_L$.

Now, we can translate the effective couplings of Eq.~\ref{lagrangian} into the exact couplings including the loop factors. In special, let us investigate how far in the $\frac{c_{BB}}{\Lambda}\times \frac{c_{WW}}{c_{BB}}$ space we are allowed to go in order to access the results from Figure.~\ref{figzw}. The scalar coupling to photons can be translated to~\cite{Franceschini:2015kwy, Altmannshofer:2015xfo}
\begin{equation}
\frac{c_{\gamma\gamma}}{\cos^2\theta_W\Lambda}\equiv\frac{c_{BB}}{\Lambda}\left(1+\frac{\sin^2\theta_W}{\cos^2\theta_W}\frac{c_{WW}}{c_{BB}}\right)=\frac{\alpha}{2\pi\cos^2\theta_W M_S}\left|N_Qd_Qq_Q^2y_Q\sqrt{\tau_Q}S(\tau_Q)+N_Lq_L^2y_L\sqrt{\tau_L}S(\tau_L)\right|
\label{Loop}
\end{equation}
where $y_Q(y_L)$ and $m_Q(m_L)$ are the heavy quark(lepton) Yukawa coupling to the scalar $S$ and the heavy quark(lepton) mass, respectively, $d_Q$ is the dimension of the $SU(3)_C$ representation, and $\tau_X=4m_X^2/M_S^2$. The loop function is given by $S(\tau)=1+(1-\tau)\arcsin^2(1/\sqrt{\tau})$ assuming $m_Q,m_L>M_S$. In the case of a  pseudoscalar $S$-cion, we should replace $y_{Q,L}\rightarrow \frac{3}{2}y_{Q,L}$ in the formula above.\footnote{Despite not being evident, $c_{\gamma\gamma}/\Lambda$ from Eq.(\ref{Loop}) scales as $1/m_F$ as expected.}

\begin{figure}[!t]
  \centering
  \includegraphics[width=0.33\textwidth]{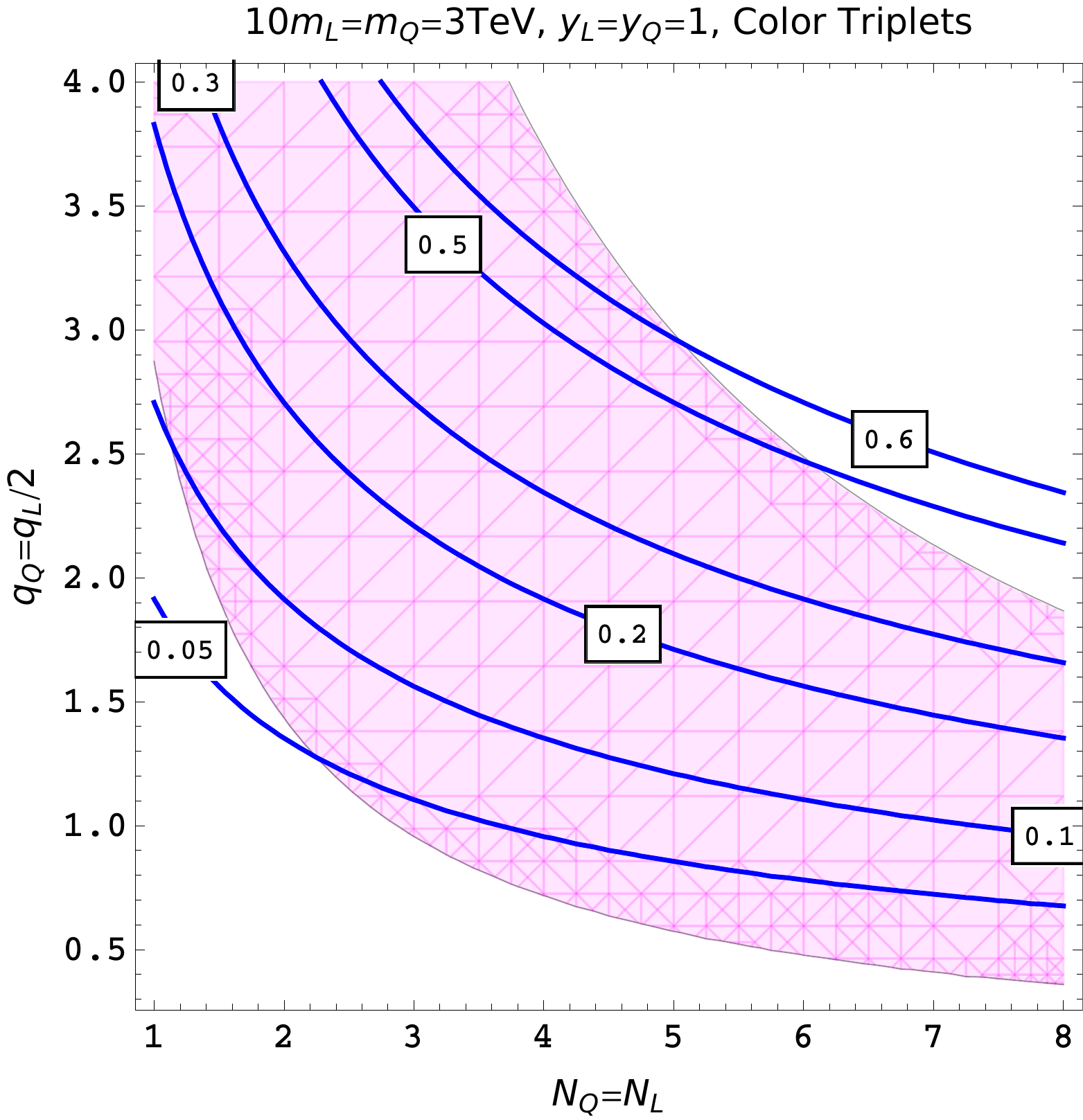}
  \includegraphics[width=0.33\textwidth]{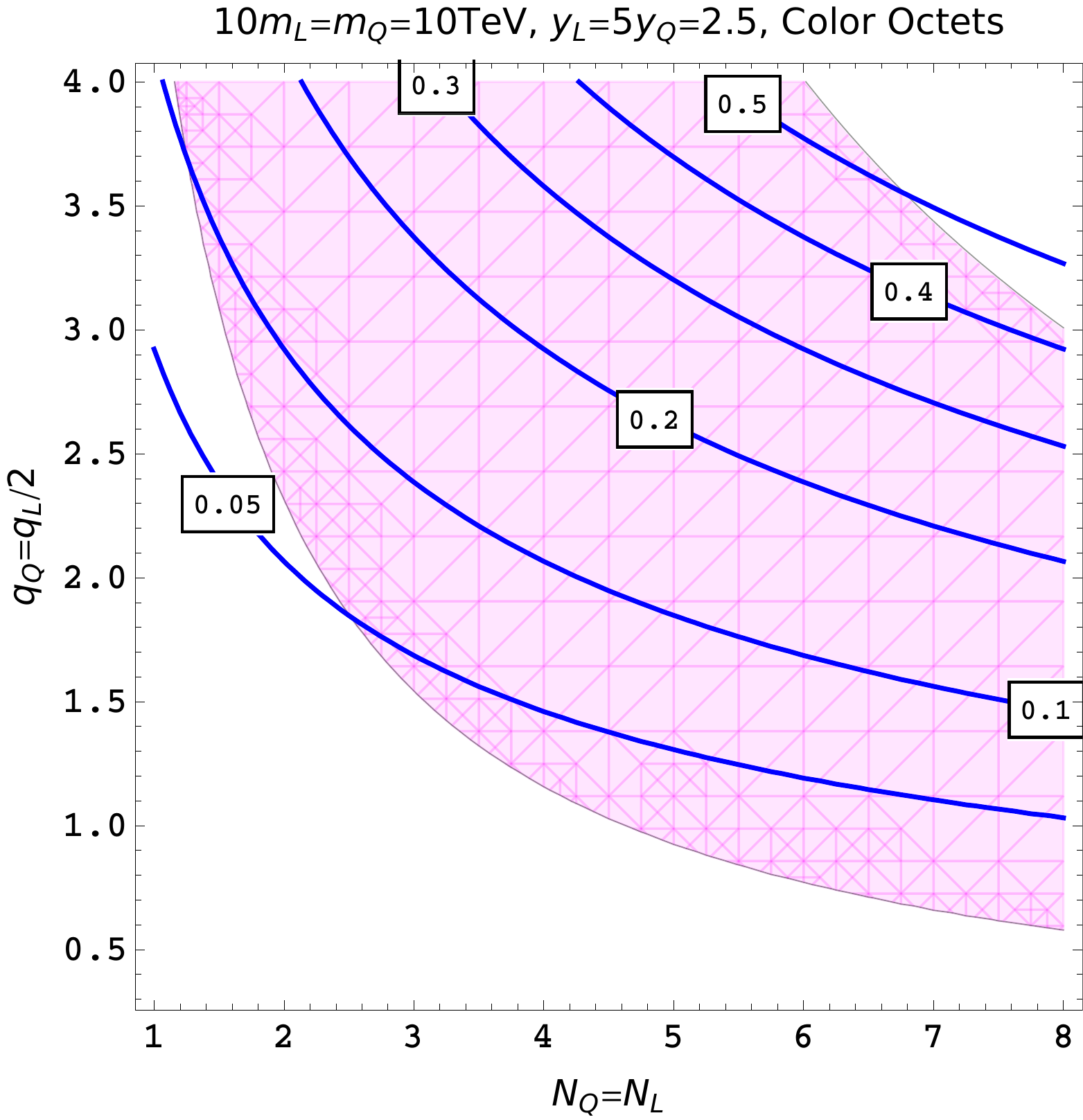}
  \includegraphics[width=0.33\textwidth]{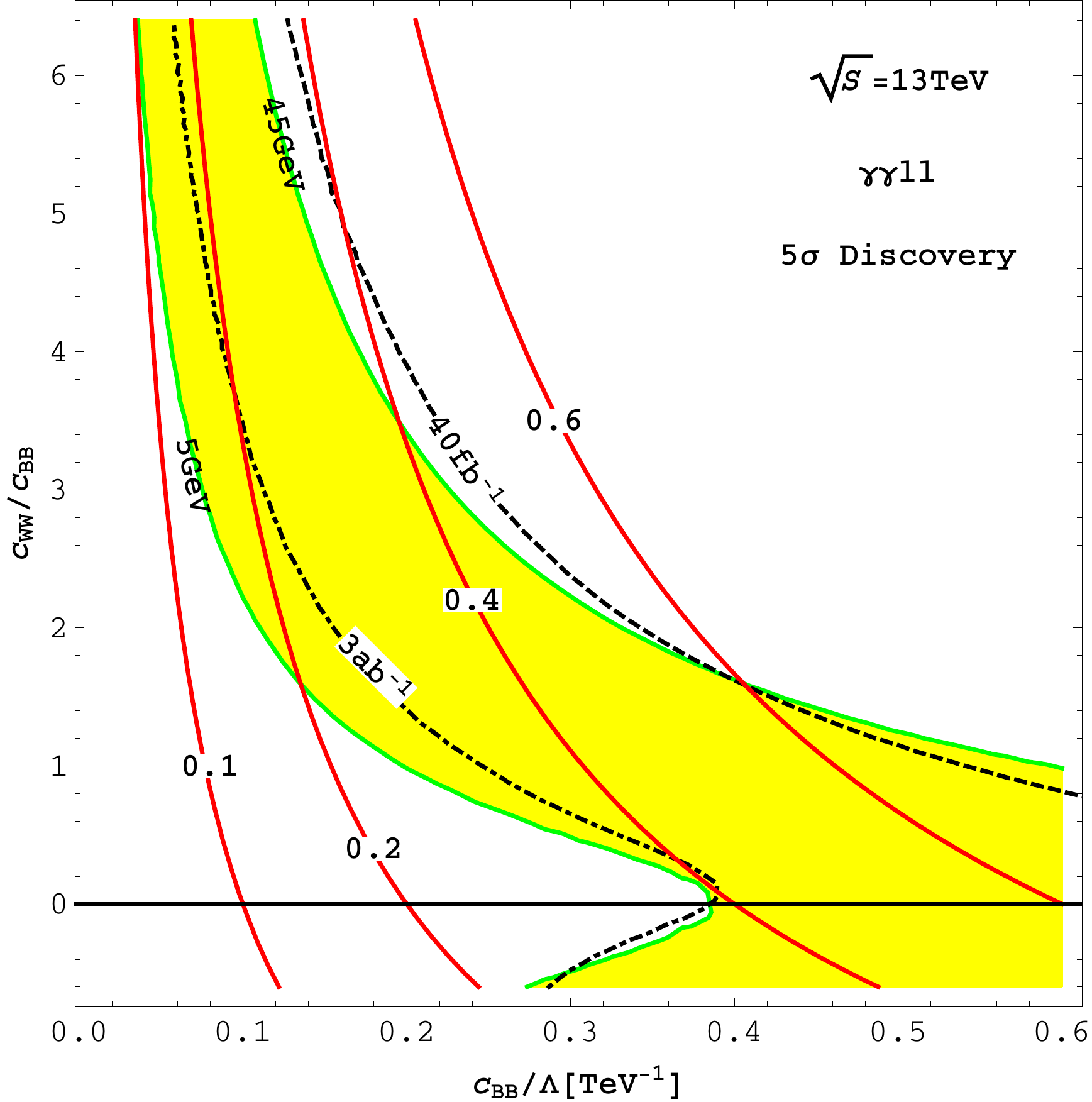}
  \caption{The lines with constant $\frac{c_{\gamma\gamma}}{\Lambda}$ (in units of TeV$^{-1}$) in the plane of models $N_Q=N_L\times y_Q=q_L/2$. Two scenarios are shown: color triplets quarks with $10m_L=m_Q=3$ TeV and $y_Q=y_L=1$ at left, and color octets with $10m_L=m_Q=10$ TeV and $5y_Q=y_L=2.5$ at the middle panel. In the right panel we show the discovery prospects of the LHC 13 TeV in the $\gamma\gamma\ell\ell$ channel for models with fixed $c_{\gamma\gamma}/\Lambda$ couplings, shown as solid red lines in the plot.}
  \label{figVL}
\end{figure}

We show, in Figure~(\ref{figVL}), the lines in the space of vector-like fermions models $N_Q=N_L\times q_Q=q_L/2$ where $c_{\gamma\gamma}/\Lambda$ is constant from 0.05 until around 0.6, assuming masses not excluded by collider searches~\cite{Khachatryan:2014mma, Chatrchyan:2013oca, Aad:2015kqa} and Yukawa couplings as displayed in the figure. The left panel shows models with color triplet quarks, and the middle panel the color octet quarks. The shaded area represents the portion of the models space where the observed diphoton excess can be fitted from the narrow resonance scenario (the lower border of the shaded area) to the wide resonance scenario (the upper border).

First of all, we observe that a $c_{\gamma\gamma}/\Lambda$ up to 0.6/TeV can be reached in scenarios with larger multiplicities and/or large electric charges within the perturbative Yukawas regime and $m_Q\geq 3$ TeV. Looking at Figure~(\ref{figzw}), we see that $c_{BB}/\Lambda\lesssim 0.5$/TeV is the bulk region of our results assuming effective couplings. The preference for those kind of models in order to fit the diphoton excess has already been pointed out in Refs.~\cite{Franceschini:2015kwy, Ellis:2015oso}, for example. Examples of models with higher quark and lepton multiplicities, higher color representations and/or exotic electric charges, and large Yukawa couplings can be found in~\cite{Sierra:2015zma, models}.

At the right panel of Figure~(\ref{figVL}), we show the discovery reach of the LHC in the $\gamma\gamma\ell\ell$ channel again, but now with solid red lines for $c_{\gamma\gamma}/\Lambda$ fixed, so we can read which points of the models space at the left and middle panels could be discovered at the LHC assuming once more the validity of the EFT parameterization. For example, let us take the intercept point where the red line (with $c_{\gamma\gamma}/\Lambda=0.6$), the dashed line (the points which can be discovered with 40 fb$^{-1}$) and the solid green line (at which the $S$-cion width is 45 GeV) cross. This point has a straightforward interpretation -- it is the point where a scalar with width 45 GeV and $c_{\gamma\gamma}/\Lambda=0.6/$TeV can be discovered with 40 fb$^{-1}$. Now, we are able to look for that point in the models space. In the triplets scenario, for example, we indeed find a point where the 0.6 line crosses the upper border of the shaded area. We can immediately read that $N_Q=N_L=5$, $q_Q=q_L/2=3$, $10m_L=m_Q=3$ TeV and Yukawa couplings $y_Q=y_L=1$. 

We could also start at some point of the models space and check if it can be discovered at the right panel. For example, in the octets scenario, with $10m_L=m_Q=10$ TeV and Yukawa couplings $5y_Q=y_L=2.5$, the point $N_Q=N_L=4$ and $q_Q=q_L/2=3$ corresponds to $c_{\gamma\gamma}/\Lambda=0.2$/TeV. We see, at the righ panel, that such a model can be discovered with 3 ab$^{-1}$ if the width is near 5 GeV and $c_{WW}\gtrsim 4c_{BB}$.


It is also possible to enhance the $S$ couplings to photons in different classes of models involving new charged scalars and/or vector bosons in the loop. For example, in 331 models, contributions from new singly and doubly charged vector bosons might enhance considerably the Higgs decays to photons~\cite{Alves:2012yp, Alves:2011kc, Yue:2013qba}. If the new physics comes with an extended gauge sector, similar situation could occur in the case of a 750 GeV scalar or pseudoscalar.

\section{Conclusions}

A new physics signal may have showed up in the ATLAS and CMS experiments suggesting the production of a 750 GeV boson, possibly a scalar or pseudoscalar particle $S$, decaying to a high mass $\gamma\gamma$ pair. This particle may be the first of a family of such particles playing important roles in the solution to the hierarchy problem.

We show in this work that both a CP-even or a CP-odd scalar singlet can produce robust new signals compatible with the LHC 8 TeV constraints and with the observed diphoton excess in very clean channels like $pp\to S+Z\to \gamma\gamma\ell\ell$, $pp\to S+W^\pm\to\gamma\gamma\ell\eslash$ and $pp\to S+W^\pm\to\ell\ell\ell\gamma\eslash$ with the current accumulated data. We parametrize the $S$ couplings to the gauge bosons in an effective field theory framework and show that a 5$\sigma$ discovery or a 3$\sigma$ evidence is possible for large portions of the parameters space. 

 Evidence for a new physics scale up to 3 TeV can already be probed in $\gamma\gamma\ell\ell$ channel with the current data where at least five events are expected on the top of very low backgrounds. With more data, however, the LHC reach extends to $\Lambda\sim 9$ TeV for up to 300 fb$^{-1}$ in all channels studied in this work in the case of a wide resonance as suggested by the ATLAS data. On the other hand, if the width turns out to be narrow, below about 20 GeV, even 3 ab$^{-1}$ will not suffice to probe the new physics in these processes.  The results are very similar for a CP-even or CP-odd $S$-cion. Moreover, we argue that searching for all the three channels is important as they probe complementary regions of the parameters space. In the absence of signals, strong constraints on the effective couplings with the weak bosons can surely be obtained.

 We also show that UV completions with non-minimal heavy vector-like quarks and leptons, for example, are expected to be observed with integrated luminosities up to 3 ab$^{-1}$ in these channels. As a matter of fact, models with enhanced decays to photons are easier to be discovered at the LHC. Models with high(low) quarks and leptons multiplicities and moderate(high) electric charges are favored in these searches. Minimal models with low multiplicities, large masses and small Yukawa couplings cannot be probed in these channels, as they are not likely to fit the diphoton excess. It is important to stress that our results apply to many other types of models not involving vector-like fermions.

The results presented in form of effective couplings, on the other hand, are important because of their extended scope. If the diphoton signal is confirmed, in a first moment where no additional information about the new physics would be available, is likely that the experimental collaborations present results in terms of effective models. Phenomenological works as this would then be important for further model building.

In resume, it is crucial to search for new signals possibly related to the diphoton excess in order to gain more insight of the type of new physics that might be arising. Other interesting channels  such as the di-Higgs and weak boson fusion should be investigated if the resonance is confirmed.


\vskip0.5cm
\textbf{Acknowledgments}
The authors acknowledge financial support
from the Brazilian agencies CNPq, under the processes 303094/2013-3
(A.G.D.), 307098/2014-1 (A.A.), FAPESP, under the process 2013/22079-8 (A.A. and A.G.D.). K.S. is supported by NASA Astrophysics Theory Grant NNH12ZDA001N.






\end{document}